\documentclass{article}

     \PassOptionsToPackage{numbers, compress}{natbib}
     \usepackage[final]{neurips_2019}

\usepackage[utf8]{inputenc} 
\usepackage[T1]{fontenc}    
\usepackage{hyperref}       
\usepackage{url}            
\usepackage{booktabs}       
\usepackage{amsfonts}       
\usepackage{nicefrac}       
\usepackage{microtype}      
\usepackage{subfigure}      
\usepackage{caption}
\usepackage{graphicx}
\usepackage{amsmath}
\usepackage{booktabs, tabularx}

\usepackage{float}

\title{DRIVE: Machine Learning to Identify Drivers of Cancer with High-Dimensional Genomic Data \& Imputed Labels}

\author{
  Adnan Akbar$^1$, Andrey Solovyev$^{1,2}$, John W Cassidy$^1$, Nirmesh Patel$^1$, Harry W Clifford$^1$ \\
  \and
  $^1$Cambridge Cancer Genomics \\
  Cambridge, UK
  \and
  $^2$The University of Edinburgh \\
  Edinburgh, UK
}

\begin{document}

\maketitle

\begin{abstract}
Identifying the mutations that drive cancer growth is key in clinical decision making and precision oncology. As driver mutations confer selective advantage and thus have an increased likelihood of occurrence, frequency-based statistical models are currently favoured. These methods are not suited to rare, low frequency, driver mutations. The alternative approach to address this is through functional-impact scores, however methods using this approach are highly prone to false positives. In this paper, we propose a novel combination method for driver mutation identification, which uses the power of both statistical modelling and functional-impact based methods. Initial results show this approach outperforms the state-of-the-art methods in terms of precision, and provides comparable performance in terms of area under receiver operating characteristic curves (AU-ROC). We believe that data-driven systems based on machine learning, such as these, will become an integral part of precision oncology in the near future.
\end{abstract}

\section{Introduction}

Cancer is a disease of the genome and is inherently heterogeneous in nature, meaning the distinctive hallmarks of a cancer can develop through a variety of mutations and biological pathways \cite{hanahan2011hallmarks}. This heterogeneity poses a unique problem when making treatment decisions compared to most other diseases. Currently, cancers are commonly treated using a generic approach based on primary tumor location rather than the underlying genomic profile. This leads to treatment failure and gives rise to resistance. This problem is compounded by the underlying genomic architecture of cancer, which is dynamic and evolves both over time and in response to therapy. A more personalized approach is required, targeting the specific genomic aberrations that confer selective advantage in an individual tumor. Identifying these \textit{"driver"} mutations is key for successful personalized treatment selection. 

In machine learning terms, differentiating driver and passenger mutations (those that are largely benign in nature) poses a standard classification problem. However, difficulty arises due to imperfect labelling and lack of clinical evidence. The idea of a \textit{"driver mutation"} is itself an imprecise concept and with no ground-truth label. Even given an arbitrary threshold for driver status, experimental validation is difficult due to the many mutations possible within a gene. In statistical approaches, frequency-based models have been the most instrumental. These are used to identify mutation rates significantly higher than observed background mutation rate \cite{brown2019finding}. However, driver mutation frequencies follow a power-law distribution - few commonly occurring mutations and a long tail of rare mutations \cite{wood2007genomic}. Detection of rare drivers is therefore limited by the amount of data available and high dimensionality of such data. Additionally, background mutation frequencies are also difficult to estimate due to variability across different samples \cite{lawrence2013mutational}. Failing to account for this leads to the increase in false positive results observed in these models. Recent studies have also explored the possibility of predicting driver mutations using functional impact-based scores \cite{gonzalez2013intogen}\cite{carter2009cancer}, however these also suffer from high false positive rates \cite{gnad2013assessment}. 

To address these issues, we provide a novel tool for driver mutation identification: DRIVE (DRiver Identification, Validation and Evaluation). DRIVE combines a statistical modelling approach with a features-based method. We computed features based on the occurrence of mutations, functional impact scores, structural properties of the genes and proteins of the associated mutations. In addition, we used ratio-metric features in our model to further improve performance. Our proposed  model is able to detect established driver mutations with higher accuracy and fewer false positives. Moreover, it provides an advantage on current models to predict rare driver mutations with higher precision. 

\section{Methods}

An overview of the different components involved in our system with data flow is shown in Figure \ref{fig:system}. A small description about each component is given below. 

\begin{figure}[H]
\centering
\includegraphics[width=0.8\columnwidth, height = 1 in]{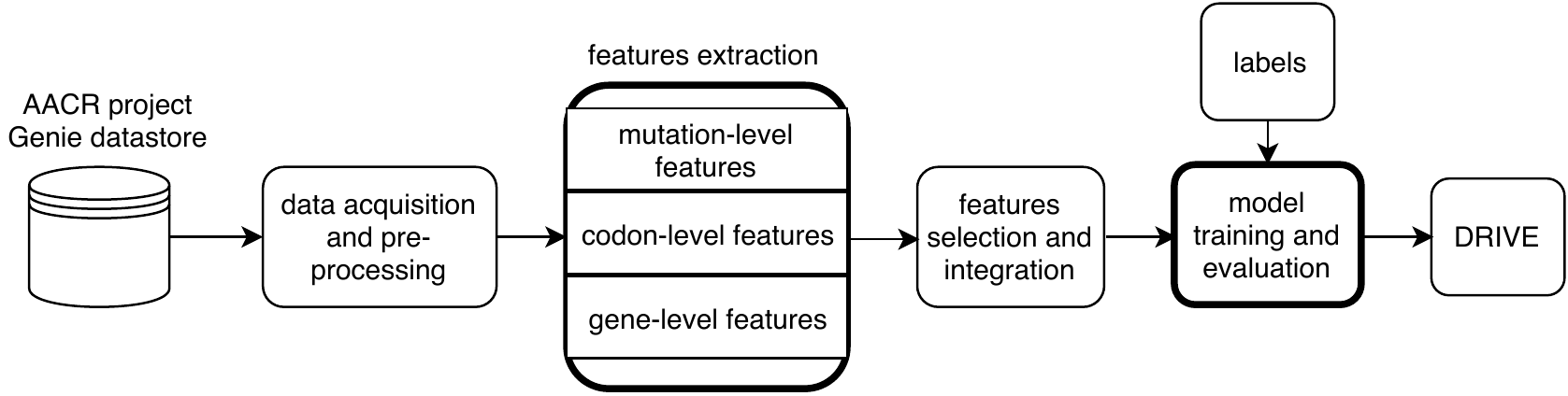}
\vspace{-2.2 mm}
\caption{Overview of DRIVE dataflow}
\label{fig:system}
\end{figure}

\subsection{Data acquisition and pre-processing}
Mutational data was obtained from the AACR Project GENIE, consisting of sequencing data of 64,217 unique tumors \cite{aacr2017aacr}. Genomic coordinates of variants from alignments to Genome Reference Consortium Human Build 37 (GRCh37) were converted to GRCh38 using PyLiftover \footnote{https://pypi.org/project/pyliftover/}. Synonymous mutations were filtered from the data, leaving 371,564 missense mutations for further analysis.

\subsection{Features extraction}
\begin{figure}[H]
\centering
\includegraphics[width=0.60\columnwidth, height = 1.65 in]{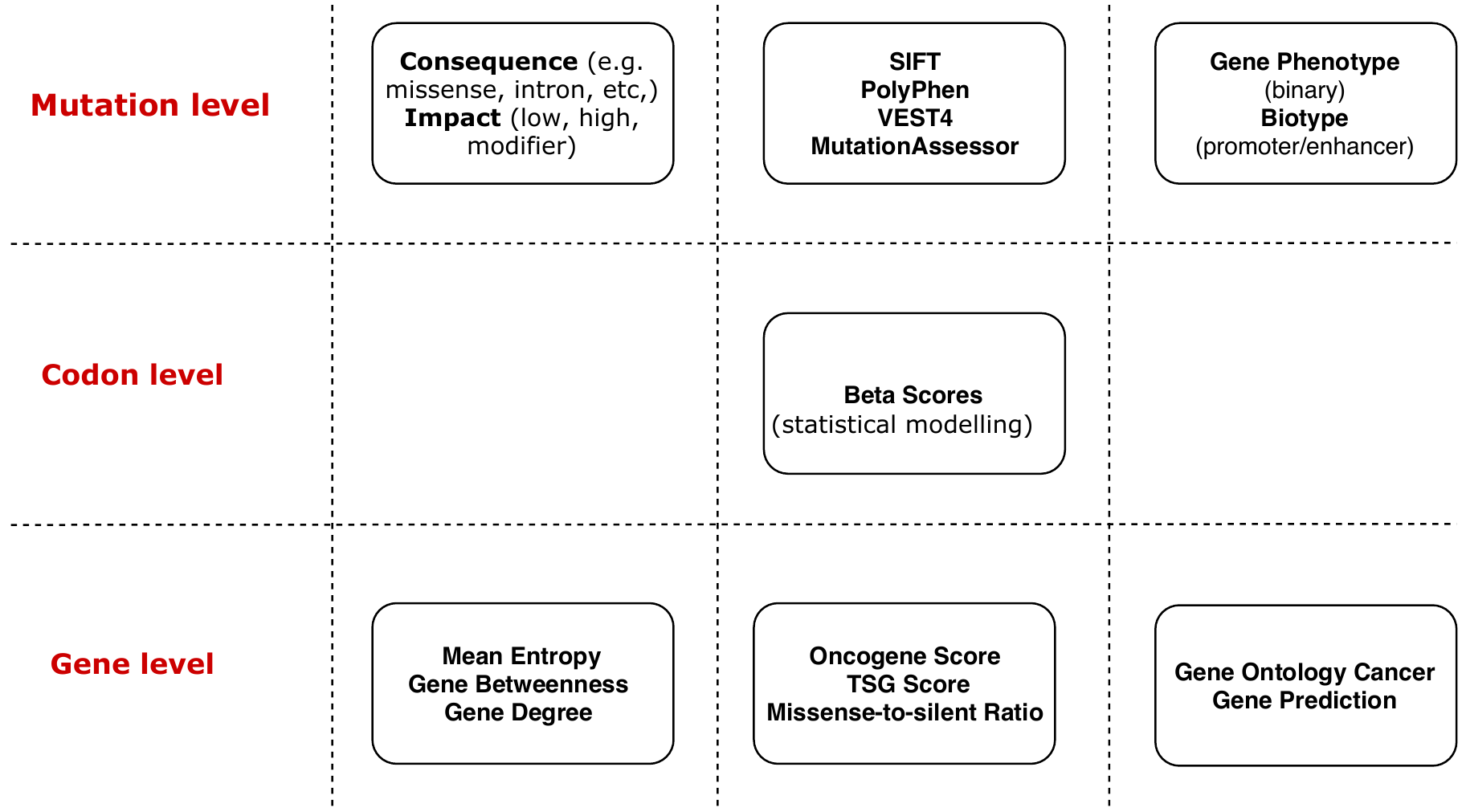}
\vspace{-2 mm}
\caption{Features used in DRIVE}
\label{fig:features}
\end{figure}

DRIVE uses a combination of features at three different levels, as shown in Figure \ref{fig:features}. Features at mutation level are computed using Variant Effect Predictor (VEP) \cite{McLaren2016} which is an open-source command line tool for analysis of genomic variants based on human genome annotations. VEP also includes features from the dbNSFP database \cite{Liu2013}, such as pathogenicity scores of Single Nucleotide Variants (SNVs) such as (SIFT \cite{Ng2003}, PolyPhen \cite{Adzhubei2010}, VEST4 \cite{Carter2013}, and MutationAssessor \cite{Reva2007}).

At codon-level, mutational hotspots were identified using the statistical modelling approach proposed in \cite{chang2016identifying}. Hotspots indicate selective pressure across a population of tumor samples and have long been regarded as the main method to identify regions with selective advantage \cite{tamborero2013comprehensive}. To counter the effect of background mutation rate, the contextual expected mutability was calculated across codons. It is used to compute the significance score of mutations known as beta scores, using a binomial model:

\begin{equation}
B_{score} = \binom{N}{k}p^k{(1-p)}^{(N-k)}
\end{equation}
where $N$ is the total number of samples sequenced, $k$ is the number of occurrence of mutations at the specific codon and $p$ is the codon mutability indicating likelihood of mutation at that locus with no selective advantage.  

Gene-level features were computed using 20/20+ \cite{Tokheim2016} - a machine learning based approach to predict driver mutations. It calculates different structural features, including Protein-protein interactions (PPI), genes degree and gene betweenness centrality. Furthermore, it calculates different ratio-metric features including non-silent to silent mutations rate and missense to silent mutations rate. 

\subsection{Data labels}
The benchmark dataset, including labels, was obtained from a recent study, ChasmPlus, where the authors used a semi-supervised approach to label the mutations as likely driver or passenger \cite{tokheim2019chasmplus}. In order for a mutation to  be labelled as a driver, it should fulfil the following three conditions; first, missense mutations must occur in the 125 clinically established pan-cancer driver gene panel \cite{vogelstein2013cancer}. Second, missense mutations of a given cancer indication must occur in a significantly mutated gene for that cancer indication \cite{lawrence2014discovery}. Third, missense mutations must be within samples with a relatively low mutation rate (less than 500 mutations) to limit the number of passenger mutations.

\subsection{Model training and evaluation}
Multiple supervised classification algorithms were implemented. We handled data imbalance with two mechanisms: first, we downsampled the majority class to equal numbers of driver and passenger mutations; second, we used stratified k-fold cross validation for model evaluation. We observed the model outputs were highly sensitive to the threshold of the classifier. For clearer insight on the performance, we evaluated the models using receiver operating characteristics (ROC) and area under the ROC (AUC-ROC).

\section{Results}
\label{results}
\begin{figure*}
\centering
\subfigure[Mean ROC curves]
{
\includegraphics[width = 0.40\columnwidth, height = 2.2in]{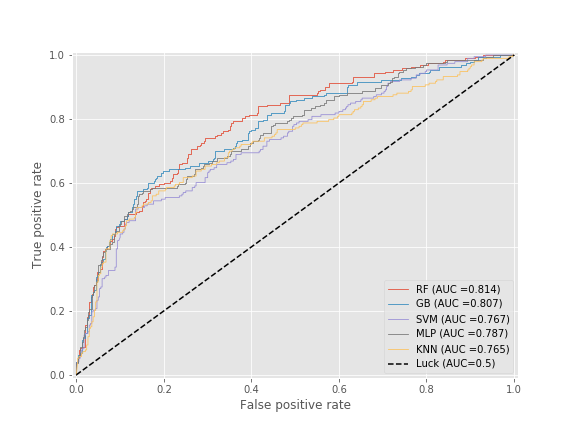}
\label{fig:roc-curves}
}
\subfigure[Random forest feature importance]
{
\includegraphics[width = 0.40\columnwidth,height = 2 in]{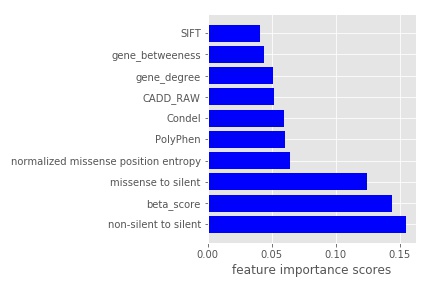}
\label{fig:second_sub}
}
\caption{Performance comparison and features importance}
\label{fig:features_importance}
\end{figure*}

Figure \ref{fig:roc-curves} displays the mean receiver operating characteristic (ROC) curves from K-fold stratified cross-validation of different classification algorithms. In general, ensemble trees such as random forest and gradient boosting classifiers performs slightly better than other models (Table \ref{table:table1}).

\begin{table}[H]
\caption{Performance evaluation using K-fold stratified cross-validation} 
\centering 
\begin{tabular}{c|c|c|c|c } 
\hline\hline 
No.&Algorithm&\texttt{ROC\_AUC}&Precision&Recall \\ [0.5ex] 
\hline 
1)&Random Forest&\textbf{0.814}&0.76&\textbf{0.60}\\
2)&Gradient Boosting&0.807&0.72&0.55\\
3)&KNN&0.765&0.75&0.53\\ 
4)&MLP&0.787&0.73&0.56\\ 
5)&SVM (ker=rbf)&0.767&\textbf{0.80}&0.33\\[1ex] 
\hline 
\end{tabular}
\label{table:table1} 
\end{table}

The values of precision and recall were computed at a threshold of 0.5, which can be adjusted to user preference. Most significant features in our system were computed using the mean decrease in Gini index (MDGI) score for random forest. Beta scores computed using the statistical model are the second most important feature in identifying driver mutations, which reiterates the advantage of our approach. Results are shown in Figure \ref{fig:features_importance}.

\begin{figure*}
\centering
\subfigure[Mean ROC curves]
{
\includegraphics[width = 0.40\columnwidth, height = 2.2 in]{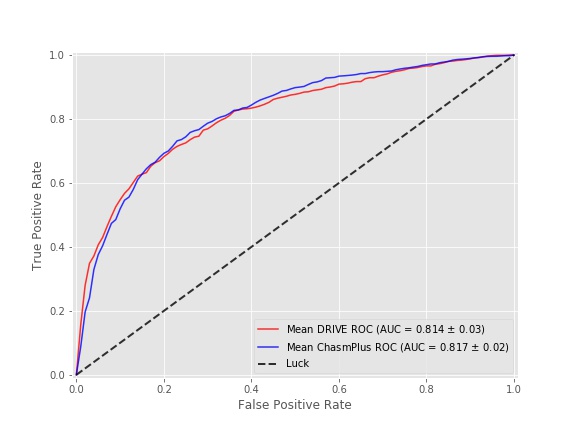}
\label{fig:first_sub}
}
\subfigure[Model comparison]
{
\includegraphics[width = 0.40\columnwidth,height = 2 in]{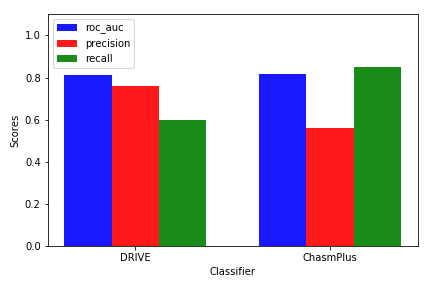}
\label{fig:second_sub}
}
\caption{Comparison with ChasmPlus \cite{tokheim2019chasmplus}}
\label{Figure:chasmplus}
\end{figure*}

We compared the performance of our highest performing classifier (random forest) with ChasmPlus \cite{tokheim2019chasmplus} (Figure \ref{Figure:chasmplus}), run in default settings \footnote{https://github.com/KarchinLab/open-cravat}. ChasmPlus scores were calculated for a pan-cancer model using OpenCRAVAT \footnote{Note: ChasmPlus has been trained on these mutations previously, which may confer an advantage}. Initial results show DRIVE performs similarly to ChasmPlus for ROC-AUC scores, and better than many currently used models (reported to have below 0.8 in \cite{tokheim2019chasmplus}). DRIVE also has better precision than ChasmPlus at a standard threshold of 0.5. 

\begin{table}[H]
\caption{Comparison with ChasmPlus} 
\centering 
\begin{tabular}{c|c|c|c|c } 
\hline\hline 
No.&Algorithm&\texttt{ROC\_AUC}&Precision&Recall \\ [0.5ex] 
\hline 
1)&DRIVE&0.814&\textbf{0.76}&0.60\\
2)&ChasmPlus&\textbf{0.817}&0.56&\textbf{0.85}\\[1ex] 
\hline 
\end{tabular}
\label{table:table2} 
\end{table}

\section{Conclusion}

Our proposed approach combines the power of statistical modelling with feature-driven machine learning methods. This provides the advantage of identifying both known cancer driver mutations and rare driver mutations with reduced false positives. We believe that these methods can be improved even further by incorporating more domain knowledge, to bring truly data-driven systems based on machine learning into precision oncology.

\bibliography{bib}

\newpage
\appendix
\section{Description of features used in DRIVE}

\begin{table}[h]
\begin{tabular}{|l|l|l|}
\hline
\multicolumn{1}{|c|}{\textbf{No.}} & \multicolumn{1}{c|}{\textbf{Feature}} & \multicolumn{1}{c|}{\textbf{Description}}                                                                                                                                              \\ \hline
1)                                 & Beta scores                           & Mutational hotspot scores using statistical model                                                                                                                                      \\ \hline
2)                                 & Missense to silent                    & Ratio of missense to silent mutations in a given gene                                                                                                                                  \\ \hline
3)                                 & Non-silent to silent                  & Ratio of non-silent to silent mutations in a given gene                                                                                                                                \\ \hline
4)                                 & HiC\_compartment                      & \begin{tabular}[c]{@{}l@{}}HiC measure of open vs consensed chromatin. Score \\ acts as a proxy for expression level\end{tabular}                                                      \\ \hline
5)                                 & Gene-betweenness                      & \begin{tabular}[c]{@{}l@{}}Betweenness centrality indicates a ratio of unique \\ paths that include a given node to all unique paths \\ in the graph.\end{tabular}                     \\ \hline
6)                                 & Gene-degree                           & \begin{tabular}[c]{@{}l@{}}Number of interaction partners on PPI network for \\ a given gene\end{tabular}                                                                              \\ \hline
7)                                 & Missense position entropy             & \begin{tabular}[c]{@{}l@{}}Mutations are binned by a codon position and for \\ each, column entropy is determined\end{tabular}                                                         \\ \hline
8)                                 & Oncogene score                        & \begin{tabular}[c]{@{}l@{}}Oncogenes are expected to have mutations clustered in\\  few positions (likely to be activating)\end{tabular}                                               \\ \hline
9)                                 & Tsg score                             & \begin{tabular}[c]{@{}l@{}}Tumour suppressor genes are expected to have \\ mutations scattered around the body of a gene \\ (likely deactivating)\end{tabular}                         \\ \hline
10)                                & Consequence                           & \begin{tabular}[c]{@{}l@{}}Sequence Ontology (SO) term that describes the effect\\  of mutation on undelying transcript (e.g. stop-gained,\\  missense, synonymous, etc.)\end{tabular} \\ \hline
11)                                & IMPACT                                & \begin{tabular}[c]{@{}l@{}}Qualitative assessment of mutation effect on the \\ function of corresponding protein (high, moderate, low)\end{tabular}                                    \\ \hline
12)                                & BIOTYPE                               & \begin{tabular}[c]{@{}l@{}}GENCODE annotation of the affected transcript. \\ Shows if mutation maps to a regulatory region,\\ transcribed region, or non-coding region.\end{tabular}   \\ \hline
13)                                & SIFT                                  & \begin{tabular}[c]{@{}l@{}}Multiple sequence alignment-based estimation of \\ tolerance to a mutation\end{tabular}                                                                     \\ \hline
14)                                & PolyPhen                              & \begin{tabular}[c]{@{}l@{}}Similar, but also incorporates the analysis of 3-D\\  structure of the affected protein\end{tabular}                                                        \\ \hline
15)                                & CADD\_Raw                             & \begin{tabular}[c]{@{}l@{}}ML-based method for distinguishing neutral mutations \\ from deleterious\end{tabular}                                                                       \\ \hline
16)                                & Condel                                & \begin{tabular}[c]{@{}l@{}}Score that integrates various computational tools for \\ consequence prediction.\end{tabular}                                                               \\ \hline
17)                                & svm\_class                            & \begin{tabular}[c]{@{}l@{}}Classification of normal genes and cancer related genes.\\ Model is based on semantic similarity between gene \\ ontology annotations\end{tabular}          \\ \hline
\end{tabular}
\end{table}
\end{document}